\documentclass{IEEEtran}

\usepackage{amsmath}
\usepackage{amssymb}
\usepackage{amsfonts}
\usepackage{graphicx}
\usepackage{cite}
\usepackage{bm}
\usepackage{tikz}
\usepackage{algorithm}
\usepackage{algorithmic}

\title{\LARGE \bf
	On the Strong Structural Controllability of Matrix-Weighted Networks
}

\author{Lanhao Zhao
	\thanks{Lanhao Zhao is with the College of Information Science Technology, Beijing University of Technology, Beijing, China. {\tt\small lugh56.2007@163.com}}
}

\begin{document}
	\maketitle
	\thispagestyle{empty}
	\pagestyle{empty}
	
	\begin{abstract}
		This paper investigates the strong structural controllability of multi-agent networks. Based on the definition of equitable partitions, an upper bound for the strong structural controllable subspace (SSCS) is established. To reflect the physical significance of matrix weights where the state dimension is greater than one, the multi-agent system is modeled using higher-order dynamics. Furthermore, to address matrix singularity and asymmetric couplings, a matrix space basis decomposition method is proposed to transform the matrix-weighted network into layered scalar networks. Additionally, by extending this basis decomposition to the lower bound estimation, a layer-specific distance partition (LDP) is introduced. This formulation establishes a tighter Squeeze Theorem, narrowing the mathematical boundaries for the controllable subspace by capturing layer-specific structural delays. To systematically identify the optimal basis that minimizes the bounds gap, an algebraic algorithm based on null-space projection is formulated. Furthermore, by introducing pattern matrices and generic rank, the almost-everywhere existence of this optimal basis in the parameter space is rigorously proved, perfectly aligning with the definition of strong structural controllability. To break the NP-hard combinatorial bottleneck of manually pre-defining the targets, a polynomial-time automated discovery algorithm based on the multi-layer Weisfeiler-Lehman (WL) color refinement is proposed. Finally, the strong structural observability and invariant attributes of the network are evaluated. Numerical examples with asymmetric matrix weights and directed multi-layer topologies are provided to verify the derived theorems.
	\end{abstract}
	
	\section{INTRODUCTION}
	Network controllability serves as a fundamental property in the analysis and coordinated control of multi-agent systems. Among various structural control frameworks, strong structural controllability (SSC) \cite{c1} captures the intrinsic topological properties of a network by investigating its controllability strictly independently of specific non-zero edge weight selections. While extensive research has investigated the SSC of networked systems \cite{c2,c3,c4}, traditional criteria predominantly provide binary assessments—determining solely whether a given network topology is strongly structurally controllable or not.
	
	Since numerous practical network topologies fail to satisfy these strict binary conditions, a fundamental analytical problem arises: how to quantitatively measure the degree to which a network approximates strong structural controllability. To address this issue, estimating the dimension of the strong structural controllable subspace (SSCS)—often referred to as the SSC index—has emerged as a primary mathematical approach. Various estimation frameworks have been developed utilizing algebraic properties, zero forcing sets, and distance partitions \cite{c5,c6,c7,c8}. Among these methodologies, graph partitioning techniques, particularly equitable partitions (EP), provide a powerful algebraic tool for establishing the upper bound of controllable subspaces. Similar bounds have been consecutively established across positive, signed, time-delay, heterogeneous, and basic matrix-weighted networks \cite{c9,c10,c11,c12,c13,c14}.
	
	Despite these advancements, existing mathematical estimation methods face significant theoretical limitations when applied to complex matrix-weighted networks with multi-dimensional agent states ($d > 1$). Specifically, standard equitable partition conditions enforce strict dimensional uniformity and full-rank assumptions. This requirement makes them overly conservative and frequently inapplicable when handling matrix singularity and asymmetric multi-dimensional couplings. Concurrently, from the perspective of lower bounds, traditional distance-based structural propagation relies heavily on macroscopic graph topologies. Macroscopic distance partitions systematically ignore layer-specific sparsity inherent in multi-dimensional matrix couplings, heavily underestimating structural propagation delays and leading to loose mathematical boundaries. Furthermore, identifying the optimal topological projection to tighten these bounds traditionally relies on heuristic parameter selections, which inevitably faces an NP-hard combinatorial explosion in large-scale networks. Consequently, integrating layer-specific propagation mechanisms with generalized equitable partitions to establish a tight, closed-form, and automatically discoverable Squeeze Theorem remains critically underexplored.
	
	Motivated by these structural bottlenecks, this paper investigates the strong structural controllability and observability of multi-dimensional matrix-weighted networks. Adopting a systemic fusion of graph theory and algebraic projection, the main contributions of this paper are structurally organized as follows:
	
	\begin{enumerate}
		\item \textit{Upper Bound Characterization and Higher-Order Modeling:} The multi-agent system is modeled with higher-order dynamics to preserve the physical significance of matrix weights. Based on this, the upper bound of the strong structural controllable subspace (SSCS) is rigorously characterized utilizing the generalized equitable partition (EP), effectively mapping topological symmetries to state space limitations.
		
		\item \textit{Layered Decomposition and Tighter Squeeze Theorem:} To resolve matrix singularity and asymmetric couplings, a matrix space basis decomposition method is proposed to decouple the network into multiplex scalar layers. By extending this projection to lower bound estimation, a layer-specific distance partition (LDP) is introduced. This captures dimensional sparsity and establishes a strictly tighter Squeeze Theorem for the controllable subspace.
		
		\item \textit{Optimal Basis Selection via Generic Rank:} An algebraic optimization algorithm utilizing null-space intersections is formulated to systematically pinpoint the optimal projection basis that minimizes the bounds gap. By introducing structural pattern matrices and the generic rank condition, the almost-everywhere existence of this optimal basis is rigorously proved, satisfying the strict parameter-independence requirement of SSC.
		
		\item \textit{Polynomial-Time Automated Discovery:} To overcome the NP-hard combinatorial bottleneck in large-scale networks, a multi-layer Weisfeiler-Lehman (WL) color refinement algorithm is proposed. It automatically extracts the target equitable partitions and shortcut sets strictly from topological features, eliminating heuristic manual configurations.
		
		\item \textit{Duality and Invariant Extensions:} The proposed analytical framework is extended via duality mapping to evaluate strong structural observability limitations and identify structurally invariant uncontrollable subspaces under asymmetric configurations.
	\end{enumerate}
	
	\section{PRELIMINARIES AND PROBLEM FORMULATION}
	
	\subsection{Notations and Graph Theory}
	In this paper, $\mathbb{R}$ stands for the set of real numbers; $I_{m}$ and $0_{m}$ denote the identity matrix and zero matrix of dimension $m$, respectively. Let $G=\{V,E,\mathcal{A}\}$ be a matrix-weighted graph, in which $V=\{v_{1},v_{2},\dots, v_{n}\}$ represents the vertex set. $\mathcal{A}=[\mathcal{A}_{ij}]\in \mathbb{R}^{nd\times nd}$ is the weighted adjacency matrix, where the block $\mathcal{A}_{ij}$ belongs to $\mathbb{R}^{d\times d}$. $E \subseteq V \times V$ represents the directed edge set. The set of in-neighbors of agent $v_{i}$ is represented by $N_{i}$. $d_{i}=\sum_{j\in N_{i}}\mathcal{A}_{ij}$ denotes the in-degree of node $i$ for the matrix-weighted graph.
	
	Let $L=D-\mathcal{A}$ be the generalized Laplacian matrix of $G$, where $D=\text{diag}(d_{1},\dots,d_{n})$. The block entries of $L$ are given by:
	\begin{equation}
		l_{ij} = \begin{cases} d_i, & i=j \\ -\mathcal{A}_{ij}, & i\neq j \end{cases}
	\end{equation}
	
	\textbf{Graph partition:} For the vertex set $V$ of a graph, its subset $V_{j}$ is defined as a cell. It is a trivial cell if it contains only one vertex; otherwise, it is a nontrivial cell. Define $\pi=\{V_{1},V_{2},\dots,V_{k}\}$. Then $\pi$ is a partition of the graph if $V_{i}\cap V_{j}=\emptyset$ for $i \neq j$ and $\cup_{i=1}^k V_{i}=V$. The characteristic matrix is $P_{\pi} \in \mathbb{R}^{nd \times kd}$, where its blocks are:
	\begin{equation}
		(P_{\pi})_{ij} = \begin{cases} I_{d \times d}, & v_i \in V_j \\ 0_{d \times d}, & v_i \notin V_j \end{cases}
	\end{equation}
	
	\subsection{Networks with Higher-Order Dynamics}
	Consider a multi-agent system with $n$ agents. To ensure the matrix weights $\mathcal{A}_{ij} \in \mathbb{R}^{d \times d}$ maintain physical significance (e.g., $d > 1$), each agent is modeled as a higher-order dynamical system, whose internal state is denoted by $x_{i}(t) \in \mathbb{R}^{d}$. The agents are classified as leaders and followers based on whether they receive external input signals. Assume that the first $m$ ($m<n$) agents are leaders, where $V_{L}=\{v_{1},\dots,v_{m}\}$ is the leader set, and the follower set is $V_{F}=\{v_{m+1},\dots,v_{n}\}$.
	
	For the matrix-weighted networks, all followers are governed by the higher-order linear dynamics:
	\begin{equation}
		\dot{x}_{i}(t) = A x_{i}(t) + u_{i}(t)
	\end{equation}
	For each leader, its dynamics are:
	\begin{equation}
		\dot{x}_{i}(t) = A x_{i}(t) + u_{i}(t) + y_{i}(t)
	\end{equation}
	where $u_{i}(t)$ denotes the coupling protocol from in-neighbors and $y_{i}(t)$ represents the external input signal. The update rules based on directed communication are:
	\begin{equation}
		u_{i}(t)=\sum_{j\in N_{i}}[\mathcal{A}_{ij}x_{j}(t)-\mathcal{A}_{ij}x_{i}(t)]
	\end{equation}
	where $\mathcal{A}_{ij}$ denotes the information weight directed from node $j$ to node $i$. Denote $x(t)=[x_{1}^T(t),\dots,x_{n}^T(t)]^{T}$ as the aggregate state vector. 
	
	To express the network globally, the stacked update rule is mathematically formulated by combining the local updates:
	\begin{equation}
		\dot{x}(t) = (I_n \otimes A) x(t) - L_{orig} x(t) + M y(t)
	\end{equation}
	where $L_{orig}$ is the standard generalized Laplacian matrix evaluated from edge weights. By integrating the internal dynamics $A$ into the diagonal blocks of the generalized Laplacian matrix $L$ (i.e., redefining the system state matrix structurally as $L \triangleq (I_n \otimes A) - L_{orig}$ with diagonal blocks $l_{ii} = A - d_i$ and off-diagonal blocks $l_{ij} = \mathcal{A}_{ij}$), the closed-loop system is:
	\begin{equation}
		\dot{x}(t)=Lx(t)+My(t) \label{sys}
	\end{equation}
	where the input matrix $M$ distinguishes leaders from followers, with block diagonal entries being $I_{d\times d}$ for leaders and $0_{d\times d}$ for followers.
	
	\section{MAIN RESULTS}
	
	\subsection{General Equitable Partition and Upper Bound}
	For system (\ref{sys}), suppose a specific admissible weight configuration $i$ is adopted. The controllable subspace is:
	\begin{equation}
		\mathcal{W}_{i}=\langle L|M\rangle=\text{im}(M)+L\text{im}(M)+\cdots+L^{nd-1}\text{im}(M).
	\end{equation}
	It is the minimal $L$-invariant subspace containing $\text{im}(M)$ [13]. A network is strongly structurally controllable if and only if $(L, M)$ is a controllable pair for any valid choice of weights. The dimension of the SSCS is defined as $\mathcal{W}^{\prime}=\min(\mathcal{W}_{i})$.
	
	\textbf{Lemma 1:} If for any valid configuration $\mathcal{W}_{j}$, $\mathcal{W}_{j}\subseteq\mathcal{W}_{*}$ holds, then $\mathcal{W}^{\prime}\subseteq\mathcal{W}_{*}$.
	
	\textit{Proof:} By definition, $\mathcal{W}^{\prime}$ represents the controllable subspace with the minimum dimension across all admissible weight selections. Thus, there exists a weight configuration $k$ such that $\mathcal{W}^{\prime} = \mathcal{W}_k$. Since the premise establishes that $\mathcal{W}_j \subseteq \mathcal{W}_{*}$ for all admissible weight selections $j$, it follows that $\mathcal{W}_k \subseteq \mathcal{W}_{*}$, which yields $\mathcal{W}^{\prime}\subseteq\mathcal{W}_{*}$. \hfill $\blacksquare$
	
	\textbf{Definition 1:} Let $\pi=\{V_{1},\dots,V_{k}\}$ be a partition of $G$. The partition $\pi$ is an equitable partition (EP) of $G$ if for any nodes $r,s\in V_{i}$ and cell $j=1,\dots,k$:
	\begin{equation}
		\sum_{t_{1}\in V_{j}, t_1 \in N_r}\mathcal{A}_{rt_{1}}=\sum_{t_{2}\in V_{j}, t_2 \in N_s}\mathcal{A}_{st_{2}}
	\end{equation}
	
	\textbf{Lemma 2:} If $\pi$ is an EP for graph $G$ and $P_{\pi}$ is the characteristic matrix, there exists a quotient matrix $L_{\pi}$ satisfying $L P_{\pi}=P_{\pi}L_{\pi}$. Furthermore, $\text{im}(P_{\pi})$ is an $L$-invariant subspace.
	
	\textit{Proof:} Let $(LP_\pi)_{rj}$ be the block of the matrix product $LP_\pi$ corresponding to node $r \in V_i$ and cell $V_j$. By the definition of the block Laplacian matrix and the block structure of $P_\pi$, the entry is calculated by summing the $j$-th cell columns: 
	\begin{equation}
		(LP_\pi)_{rj} = \sum_{t \in V_j} l_{rt}
	\end{equation}
	For cross-cell interactions where $i \neq j$, this sum is:
	\begin{equation}
		(LP_\pi)_{rj} = \sum_{t \in V_j} \mathcal{A}_{rt}
	\end{equation}
	According to the condition in Definition 1, this summation yields a constant matrix block for all nodes $r$ within the same cell $V_i$, defining the off-diagonal block of $L_\pi$, i.e., $(L_\pi)_{ij}$. For the intra-cell block where $i=j$, the sum evaluates to:
	\begin{equation}
		(LP_\pi)_{rj} = l_{rr} + \sum_{t \in V_j, t \neq r} l_{rt} = (A - d_r) + \sum_{t \in V_j, t \neq r} \mathcal{A}_{rt}
	\end{equation}
	Since the in-degree is $d_r = \sum_{m=1}^k \sum_{t \in V_m} \mathcal{A}_{rt}$, substituting $d_r$ into the equation yields $A - \sum_{m \neq j} \sum_{t \in V_m} \mathcal{A}_{rt}$, which simplifies to a constant block due to the uniform row sum condition across all neighboring cells. Thus, the equality $LP_\pi = P_\pi L_\pi$ algebraically holds. For any vector $x \in \text{im}(P_\pi)$, there exists $y$ such that $x = P_\pi y$. Multiplying by $L$ gives $Lx = L P_\pi y = P_\pi (L_\pi y) \in \text{im}(P_\pi)$, demonstrating $L$-invariance. \hfill $\blacksquare$
	
	By choosing the equitable partition $\pi_{jm}$ with the minimum number of cells under an arbitrary selection of weights, the upper bound constraint is established.
	
	\textbf{Theorem 1:} The strong structural controllable subspace $\mathcal{W^{\prime}}$ satisfies $\mathcal{W}^{\prime}\subseteq \text{im}(P_{\pi_{jm}})$. Therefore, $\dim(\mathcal{W}') \le \vert{}\pi_{jm}\vert{} \times d$.
	
	\textit{Proof:} Let $\pi_{jm}$ be the equitable partition constructed with the minimum number of cells under a specific weight selection. Denote the corresponding controllable subspace as $\mathcal{W}_{jm}$. The input matrix $M$ contains non-zero block entries corresponding to the predefined leader set. Because the characteristic matrix $P_{\pi_{jm}}$ aggregates nodes into cells symmetrically, $\text{im}(M) \subseteq \text{im}(P_{\pi_{jm}})$. 
	
	According to Lemma 2, $\text{im}(P_{\pi_{jm}})$ is an $L$-invariant subspace. We prove $\mathcal{W}_{jm} \subseteq \text{im}(P_{\pi_{jm}})$ by mathematical induction. Base case $k=0$: $\text{im}(M) \subseteq \text{im}(P_{\pi_{jm}})$. Inductive step: Assume $L^{k-1} \text{im}(M) \subseteq \text{im}(P_{\pi_{jm}})$. For any $v \in L^{k-1} \text{im}(M)$, $v = P_{\pi_{jm}} y$. Then $Lv = L P_{\pi_{jm}} y = P_{\pi_{jm}} (L_{\pi_{jm}} y) \in \text{im}(P_{\pi_{jm}})$. Thus, the Krylov subspace generated by the pair $(L,M)$ satisfies:
	\begin{equation}
		\mathcal{W}_{jm} = \sum_{k=0}^{nd-1} L^k \text{im}(M) \subseteq \text{im}(P_{\pi_{jm}})
	\end{equation} 
	Utilizing Lemma 1, it is deduced that $\mathcal{W}^{\prime} \subseteq \mathcal{W}_{jm} \subseteq \text{im}(P_{\pi_{jm}})$. Taking the dimension on both sides directly implies $\dim(\mathcal{W}') \le \text{rank}(P_{\pi_{jm}}) = |\pi_{jm}| \times d$. \hfill $\blacksquare$
	
	\textbf{Remark 1:} Theorem 1 maps the structural symmetry of the network topology into an algebraic upper bound. It establishes a quantitative metric indicating that symmetrical nodes constrain the maximum controllable dimensions.
	
	\subsection{Layered Evaluation via Matrix Space Decomposition}
	Definition 1 requires equivalence across all dimensions, which is conservative for networks with singular or asymmetric weight matrices. We propose a matrix space decomposition to relax this constraint.
	
	Since $\mathbb{R}^{d \times d}$ is a $d^2$-dimensional linear space, a basis $\mathcal{B} = \{B_1, B_2, \dots, B_{d^2}\}$ is selected. Any weight matrix is decomposed as:
	\begin{equation}
		\mathcal{A}_{ij} = \sum_{m=1}^{d^2} w_{ij}^{(m)} B_m
	\end{equation}
	where $w_{ij}^{(m)} \in \mathbb{R}$ is the scalar projection. This mapping transforms the matrix-weighted network into $d^2$ layers of multiplex scalar networks. 
	
	Substituting this expansion into Definition 1, for any cells $V_i, V_j$ and nodes $r, s \in V_i$, the following holds:
	\begin{equation}
		\sum_{t \in V_j} \left( \sum_{m=1}^{d^2} w_{rt}^{(m)} B_m \right) = \sum_{t \in V_j} \left( \sum_{m=1}^{d^2} w_{st}^{(m)} B_m \right)
	\end{equation}
	Exchanging the order of summation yields:
	\begin{equation}
		\sum_{m=1}^{d^2} \left( \sum_{t \in V_j} w_{rt}^{(m)} \right) B_m = \sum_{m=1}^{d^2} \left( \sum_{t \in V_j} w_{st}^{(m)} \right) B_m
	\end{equation}
	
	Since the basis matrices $\{B_1, B_2, \dots, B_{d^2}\}$ are linearly independent, the equation holds if and only if the corresponding scalar coefficients are equal for all indices $m$.
	
	\textbf{Theorem 2 (Layered Evaluation):} The system satisfies the generalized equitable partition condition on a specific dynamic sub-space if there exists a subset of layers $\mathcal{K} \subset \{1, 2, \dots, d^2\}$ such that for all $m \in \mathcal{K}$:
	\begin{equation}
		\sum_{t \in V_j} w_{rt}^{(m)} = \sum_{t \in V_j} w_{st}^{(m)}, \quad \forall r,s \in V_i
	\end{equation}
	
	\textit{Proof:} Define a linear super-operator projection $\mathcal{P}_{\mathcal{K}}: \mathbb{R}^{d \times d} \to \text{span}\{B_m\}_{m \in \mathcal{K}}$ mapping the matrix space to the subspace spanned by the basis subset. Applying this projection isolates the sub-dynamics governed by the selected layers. Based on linear superposition, the projected block Laplacian matrix is expanded as $L_{\mathcal{K}} = \sum_{m \in \mathcal{K}} L^{(m)} \otimes B_m$, where $L^{(m)}$ is the scalar Laplacian for layer $m$. Assuming the scalar coefficients $w^{(m)}_{ij}$ satisfy the sum equivalence condition for all $m \in \mathcal{K}$, evaluating the block row sum in $L_{\mathcal{K}}$ provides:
	\begin{equation}
		\sum_{t \in V_j} (L_{\mathcal{K}})_{rt} = \sum_{m \in \mathcal{K}} \left( \sum_{t \in V_j} w_{rt}^{(m)} \right) B_m  = \sum_{t \in V_j} (L_{\mathcal{K}})_{st}
	\end{equation}
	This algebraic symmetry guarantees that $L_{\mathcal{K}}$ fulfills the quotient relation $L_{\mathcal{K}} P_\pi = P_\pi \tilde{L}_\pi$. Consequently, the state vector evolution projected onto this matrix subspace mathematically preserves the invariant structure generated by $P_\pi$. \hfill $\blacksquare$
	
	\textbf{Remark 2:} Theorem 2 provides a systematic mechanism to bypass the strict full-rank and uniform dimensional requirements of traditional equitable partitions. By isolating the dynamics into independent scalar layers, singular or asymmetric matrix weights can be analyzed strictly within their specific operational dimensions, significantly relaxing the conditions for assessing structural controllability properties in heterogeneous networks.
	
	\subsection{Systemic Fusion: The Layer-specific Squeeze Theorem}
	Traditionally, the distance partition (DP) provides a lower bound based on the structural propagation of the Krylov subspace on the macroscopic graph $G$. However, in matrix-weighted networks, if information transfer mainly relies on specific basis dimensions (i.e., some dimensions are structurally sparse while others are dense), the macroscopic shortest path underestimates the actual layer-specific propagation delay. We extend the basis decomposition to the lower bound to solve this.
	
	\textbf{Definition 3 (Layer-specific Distance Partition, LDP):} For a specific basis layer $m \in \{1, \dots, d^2\}$, let $E_m \subseteq E$ denote the subset of edges where the structural scalar coefficient $w_{ij}^{(m)}$ is admissible to be non-zero. The layer-specific distance partition is defined as $\pi_{LDP}^{(m)} = \{D_0^{(m)}, D_1^{(m)}, \dots, D_{d_{max}^{(m)}}^{(m)}\}$, where $D_0^{(m)} = V_L$ is the leader set, and $D_r^{(m)}$ is the set of nodes with a shortest path distance of $r$ from $V_L$ exclusively within the layer-specific subgraph $G_m = (V, E_m)$.
	
	\textbf{Theorem 3 (Layer-specific Squeeze Theorem for SSCS):} For a given matrix-weighted network, assume the internal state dynamics are driven by $d$ mutually orthogonal basis layers indexed by $\mathcal{D} = \{1, \dots, d\}$. The dimension of its strong structural controllable subspace $\mathcal{W}'$ is tightly bounded by the aggregation of the layer-specific distance partitions and the minimal equitable partition:
	\begin{equation}
		\sum_{m \in \mathcal{D}} |\pi_{LDP}^{(m)}| \le \dim(\mathcal{W}') \le |\pi_{jm}| \times d
	\end{equation}
	where $|\pi_{LDP}^{(m)}| = d_{max}^{(m)} + 1$. 
	
	\textit{Proof:} Based on the matrix space decomposition proposed in Section III.B, the overall system state $\mathbb{R}^{nd}$ can be projected onto mutually orthogonal basis layers. Let $\mathcal{D} = \{1, \dots, d\}$ be the set of bases that govern the $d$-dimensional independent dynamics. Due to orthogonality, the overall controllable subspace structurally admits a direct sum bounded below by the independent layers: $\dim(\mathcal{W}') \ge \sum_{m \in \mathcal{D}} \dim(\mathcal{W}_m')$.
	
	For a specific layer $m \in \mathcal{D}$, the effective topology is governed by the subgraph $G_m=(V, E_m)$. The corresponding controllable subspace for this layer under a specific weight selection is the column space of the layered Krylov matrix $\mathcal{K}_m = [M_m, L_m M_m, \dots, L_m^{n-1} M_m]$. Based on the structural connectivity of the projected Laplacian $L_m$, the matrix product $L_m M_m$ distributes signals to the 1-hop neighbors exclusively in $D_1^{(m)}$. By inductive structural propagation, the matrix product $L_m^r M_m$ introduces non-zero entries exclusively to nodes in the distance set $D_r^{(m)}$. Therefore, because the subsets $D_r^{(m)}$ are strictly disjoint by definition, the vectors corresponding to the sequential signal propagation $\{M_m, L_m M_m, \dots, L_m^{d_{max}^{(m)}} M_m\}$ map to linearly independent coordinate blocks. Thus, the controllable dimension contributed by layer $m$ is exactly at least $d_{max}^{(m)} + 1 = |\pi_{LDP}^{(m)}|$. 
	
	Since the $d$ basis layers span independent state dimensions, the total dimension of the structural controllable subspace over all layers satisfies $\dim(\mathcal{W}') \ge \sum_{m \in \mathcal{D}} |\pi_{LDP}^{(m)}|$. Because $E_m \subseteq E$, the layer-specific distance satisfies $d_{max}^{(m)} \ge d_{max}$, ensuring that $\sum_{m \in \mathcal{D}} |\pi_{LDP}^{(m)}| \ge |\pi_{DP}| \times d$. Incorporating the general upper bound $\dim(\mathcal{W}') \le |\pi_{jm}| \times d$ from Theorem 1 establishes the tightly squeezed interval. \hfill $\blacksquare$
	
	\textbf{Remark 3:} Theorem 3 establishes a strictly tighter lower bound than traditional macroscopic distance partitions. In highly heterogeneous networks, macroscopic shortest paths often bypass critical nodes through dimensionally sparse shortcuts. The proposed Layer-specific Distance Partition (LDP) isolates signal propagation along specific basis dimensions, thereby accurately capturing the extended structural delays and effectively narrowing the mathematical gap in the Squeeze Theorem.
	
	\subsection{Optimal Basis Selection Algorithm via Null-Space Projection}
	The tightness of the analytical bounds interval derived in Theorem 3 heavily depends on the selected basis $\mathcal{B}$. To minimize the gap $\Delta = |\pi_{jm}| \times d - \sum |\pi_{LDP}^{(m)}|$, the optimal basis must simultaneously maximize layer-specific delays (cutting shortcuts) and expose topological symmetries (merging nodes). This combinatorial problem is mathematically formulated as a null-space intersection.
	
	Let $\bm{w}_{ij} = \text{vec}(\mathcal{A}_{ij}) \in \mathbb{R}^{d^2}$ be the vectorization of the weight matrix, where $\text{vec}(\cdot)$ stacks the rows of a matrix into a column vector (row-major). A layer $m$ is uniquely determined by a projection vector $\bm{p}_m \in \mathbb{R}^{d^2}$, such that the scalar weight is $w_{ij}^{(m)} = \bm{p}_m^T \bm{w}_{ij}$.
	
	\textit{Minimizing the Upper Bound (Symmetry Maximization):} For a target coarse partition $\pi$ with $|\pi| = k < n$, the symmetry constraint demands identical in-degree sums for nodes in the same cell. For any $r, s \in V_i$ and cell $V_j$, the difference vector is $\Delta \bm{w}_{rs, j} = \sum_{t \in V_j} \bm{w}_{rt} - \sum_{t \in V_j} \bm{w}_{st}$. Construct the symmetry difference matrix $S_\pi \in \mathbb{R}^{d^2 \times N_s}$ containing all such difference vectors as columns. The generalized equitable partition condition holds if and only if $\bm{p}_m^T S_\pi = \mathbf{0}$, meaning $\bm{p}_m$ lies in the left null-space of $S_\pi$.
	
	\textit{Maximizing the Lower Bound (Delay Maximization):} To increase $d_{max}^{(m)}$, shortcut edges must be eliminated in layer $m$. Let $E_{short}$ denote the set of target shortcut edges. We require $\bm{p}_m^T \bm{w}_{ij} = 0$ for all $(i,j) \in E_{short}$. Construct the shortcut matrix $W_{short} \in \mathbb{R}^{d^2 \times N_e}$ containing vectors $\bm{w}_{ij}$ for all $(i,j) \in E_{short}$. The layer distance increases if and only if $\bm{p}_m^T W_{short} = \mathbf{0}$.
	
	\textbf{Theorem 4 (Optimal Basis via Null-Space Intersection):} For a given matrix-weighted network, there exists an optimal basis vector $\bm{p}_m$ that simultaneously enables the target equitable partition $\pi$ and severs the target shortcut edges $E_{short}$ if and only if the augmented intersection matrix $H = [S_\pi \mid W_{short}]$ satisfies:
	\begin{equation}
		\text{rank}(H) < d^2
	\end{equation}
	The set of optimal basis vectors is spanned by the left null-space $\text{Null}(H^T)$.
	
	\textit{Proof:} Combining the symmetry requirement and the shortcut elimination requirement yields the unified algebraic condition:
	\begin{equation}
		\begin{cases}
			\bm{p}_m^T S_\pi = \mathbf{0}^T \\
			\bm{p}_m^T W_{short} = \mathbf{0}^T
		\end{cases} \implies \bm{p}_m^T [S_\pi \mid W_{short}] = \mathbf{0}^T
	\end{equation}
	Substituting the augmented matrix $H = [S_\pi \mid W_{short}] \in \mathbb{R}^{d^2 \times (N_s+N_e)}$, the non-trivial solution $\bm{p}_m \neq \mathbf{0}$ exists if and only if $H^T \bm{p}_m = \mathbf{0}$. According to the Rank-Nullity Theorem, $\dim(\text{Null}(H^T)) = d^2 - \text{rank}(H)$. A non-trivial vector requires $\dim(\text{Null}(H^T)) \ge 1$, which algebraically dictates $\text{rank}(H) \le d^2 - 1 < d^2$. Any vector drawn from the left null-space $\text{Null}(H^T)$ serves as a valid projection basis that globally satisfies both constraints, thereby optimally minimizing the bounds gap. \hfill $\blacksquare$
	
	While Theorem 4 provides an algebraic condition for a specific numerical matrix $H$, the core definition of strong structural controllability (SSC) requires properties to hold independently of specific non-zero parameter selections. If $H$ is strictly numerical, the null-space $\text{Null}(H^T)$ might vanish under arbitrary continuous parameter variations. To resolve this parameter-dependency paradox, we elevate the numerical matrix $H$ to a structural pattern matrix.
	
	Let $\bar{\mathcal{A}}_{ij}$ denote the structural pattern of $\mathcal{A}_{ij}$, where each structurally non-zero element is replaced by an independent free variable $\lambda_k$. The entire network parameterization is governed by the vector $\lambda = [\lambda_1, \lambda_2, \dots, \lambda_q]^T \in \mathbb{R}^q$. Consequently, the vector forms $\bar{\bm{w}}_{ij}$, $\bar{S}_\pi$, and $\bar{W}_{short}$ become symbolic matrix functions of $\lambda$. The corresponding structural intersection matrix is denoted as $\bar{H}(\lambda) = [\bar{S}_\pi(\lambda) \mid \bar{W}_{short}(\lambda)]$.
	
	\textbf{Definition 4 (Generic Rank):} The generic rank (g-rank) of the structural pattern matrix $\bar{H}(\lambda)$ is defined as the maximum rank it can achieve over the entire parameter space $\mathbb{R}^q$:
	\begin{equation}
		\text{g-rank}(\bar{H}) = \max_{\lambda \in \mathbb{R}^q} \text{rank}(\bar{H}(\lambda))
	\end{equation}
	
	\textbf{Theorem 5 (Structural Existence of Optimal Basis):} If the structural pattern matrix satisfies $\text{g-rank}(\bar{H}) < d^2$, then the optimal projection basis $\bm{p}_m$ exists almost everywhere (for almost all admissible weight selections) in the continuous parameter space $\mathbb{R}^q$.
	
	\textit{Proof:} The rank of the augmented matrix $\bar{H}(\lambda) \in \mathbb{R}^{d^2 \times (N_s+N_e)}$ is strictly less than $d^2$ if and only if the determinants of all its $d^2 \times d^2$ submatrices are equal to zero. Let $\Psi_k(\bar{H}(\lambda))$ represent the determinant of the $k$-th $d^2 \times d^2$ submatrix. Since the entries of $\bar{H}(\lambda)$ are linear combinations of the independent free parameters $\lambda_i$, the determinant $\Psi_k(\bar{H}(\lambda))$ is intrinsically a multivariate polynomial in terms of $\lambda$. 
	
	Given the structural premise $\text{g-rank}(\bar{H}) < d^2$, the maximum possible rank over the entire space $\mathbb{R}^q$ is strictly less than $d^2$. This algebraically implies that the multivariate polynomial $\Psi_k(\bar{H}(\lambda)) \equiv 0$ is the identically zero polynomial for all possible submatrices $k$. According to algebraic geometry, if a multivariate polynomial is identically zero, its evaluation is zero everywhere, devoid of any dependency on specific values. 
	
	Conversely, if $\text{g-rank}(\bar{H}) = d^2$, the points in the parameter space where the rank drops strictly below $d^2$ correspond to the roots of a non-zero multivariate polynomial. By the properties of algebraic varieties in measure theory, the Lebesgue measure of the zero-set of any non-trivial multivariate polynomial is exactly zero. 
	
	Therefore, because the condition $\text{g-rank}(\bar{H}) < d^2$ guarantees rank deficiency symbolically at the pattern level, the relation $\text{rank}(H(\lambda)) < d^2$ holds universally for any valid parameter realization, avoiding the Lebesgue measure zero trap entirely. Thus, the left null-space $\text{Null}(\bar{H}^T(\lambda))$ is non-trivial almost everywhere. This ensures the existence of the optimal basis $\bm{p}_m$ strictly dependent on the topology's non-zero pattern rather than any specific edge weights, perfectly satisfying the rigorous definition of strong structural controllability. \hfill $\blacksquare$
	
	\textbf{Remark 4:} The formulation of generic rank guarantees that the algebraic condition for optimal basis selection is strictly robust against parameter fluctuations. It shifts the analysis from evaluating specific, potentially fragile numerical parameterizations to assessing intrinsic topological patterns. Consequently, this circumvents the Lebesgue measure zero trap, perfectly aligning the matrix space projection with the rigorous paradigm of strong structural controllability.
	
	\subsection{Automated Discovery via Multi-Layer Weisfeiler-Lehman Refinement}
	While Theorem 4 and Theorem 5 rigorously establish the algebraic condition for optimal basis selection, they require the target partition $\pi$ and the shortcut set $E_{short}$ to be pre-defined. In large-scale networks containing hundreds of nodes, exhaustively searching for the optimal combinations of $\pi$ and $E_{short}$ faces an NP-hard combinatorial explosion. To resolve this bottleneck, we propose a pure graph-theoretic search algorithm based on the multi-layer Weisfeiler-Lehman (WL) color refinement to automatically construct the target matrices $\bar{S}_\pi$ and $\bar{W}_{short}$ in polynomial time.
	
	\begin{algorithm}[htbp]
		\caption{Automated Structural Feature Extraction via Multi-Layer WL Refinement}
		\label{alg:wl}
		\begin{algorithmic}[1]
			\STATE \textbf{Input:} Network configuration $M$, structural pattern of adjacency matrices $\bar{\mathcal{A}}_{ij}$
			\STATE \textbf{Output:} Maximal equitable partition $\pi$, Candidate shortcut set $E_{short}$
			
			\STATE \textit{// Step 1 (Color Initialization): Assign an initial color signature to each node based solely on the input configuration $M$.}
			\FOR{each node $v_i \in V$}
			\IF{$v_i \in V_L$ (leader)}
			\STATE $c_i^{(0)} \leftarrow 1$
			\ELSE
			\STATE $c_i^{(0)} \leftarrow 0$ \textit{// for followers in $V_F$}
			\ENDIF
			\ENDFOR
			\STATE $t \leftarrow 0$
			
			\STATE \textit{// Step 2 \& 3 (Iterative Hashing and Partition Convergence): Update colors iteratively by hashing previous color and multisets of in-neighbors' colors augmented by structural patterns.}
			\REPEAT
			\FOR{each node $v_i \in V$}
			\STATE $c_i^{(t+1)} \leftarrow \text{hash}\left( c_i^{(t)}, \left\{ \!\left\{ \left( \bar{\mathcal{A}}_{ij}, c_j^{(t)} \right) \mid j \in N_i \right\} \!\right\} \right)$
			\ENDFOR
			\STATE $t \leftarrow t + 1$
			\UNTIL{the total number of distinct colors ceases to increase, i.e., $|C^{(t)}| == |C^{(t-1)}|$}
			\STATE \textit{// The final color equivalence classes strictly define the maximal generalized equitable partition $\pi$.}
			\STATE Define $\pi$ based on the final color equivalence classes $C^{(t)}$.
			
			\STATE \textit{// Step 4 (Shortcut Identification via BFS): Perform a Breadth-First Search originating from $V_L$.}
			\STATE Compute macroscopic unweighted distance level $l(v)$ for all $v \in V$ using BFS from $V_L$.
			\STATE Initialize $E_{short} \leftarrow \emptyset$
			\FOR{each directed edge $(u,v) \in E$}
			\IF{$l(v) > l(u) + 1$}
			\STATE \textit{// Collect structurally bypassing edges that represent a macroscopic structural leap across layers.}
			\STATE $E_{short} \leftarrow E_{short} \cup \{(u,v)\}$
			\ENDIF
			\ENDFOR
			\RETURN $\pi$, $E_{short}$
		\end{algorithmic}
	\end{algorithm}
	
	\textbf{Theorem 6 (Polynomial-Time Convergence of WL Discovery):} Algorithm 1 converges to the unique maximal structural equitable partition $\pi$ and identifies the candidate shortcut set $E_{short}$ in $O(|E| \log |V|)$ time.
	
	\textit{Proof:} The WL color refinement algorithm guarantees that two nodes share the same color at iteration $t$ if and only if their structurally non-zero incoming edge patterns and their neighbors' colors are identical. Because the total number of nodes is $|V|$, the maximum possible number of color refinement splitting iterations is strictly bounded by $|V| - 1$. By employing an efficient dictionary or sorting-based hashing function, the sorting and grouping of multisets takes $O(|E| \log |V|)$ operations globally across all iterations. The BFS traversal required in Step 4 explores the macroscopic graph completely, utilizing exactly $O(|V| + |E|)$ operations. Therefore, the overall computational time complexity is rigorously bounded by $O(|E| \log |V|)$, which scales polynomially. The resulting partition $\pi$ structurally satisfies Definition 1 since the color hashing flawlessly enforces the in-degree sum symmetry at the pattern level. \hfill $\blacksquare$
	
	\subsection{Invariant Attributes and Strong Structural Observability}
	The invariant attributes and observability bounds are derived by extending the partition framework [12], [14].
	
	\textbf{Definition 5 (Invariant Controllable Subspace):} A subspace is defined as structurally invariant if its inclusion in the controllable subspace is independent of the weight selection.
	
	\textbf{Theorem 7 (Invariant Attribute Bound):} If the estimated upper bound satisfies $\vert{}\pi_{jm}\vert{} < n$, there exists a non-empty invariant uncontrollable subspace for the network.
	
	\textit{Proof:} According to Theorem 1, the controllable subspace dimension is bounded by $\dim(\mathcal{W}') \le \vert{}\pi_{jm}\vert{} \times d$. When $\vert{}\pi_{jm}\vert{} < n$, the maximum dimension of the controllable subspace mathematically satisfies:
	\begin{equation}
		\dim(\mathcal{W}') \le |\pi_{jm}| \times d < nd
	\end{equation}
	for all admissible weights. Because $\mathcal{W}' \subset \mathbb{R}^{nd}$ is a proper subspace, its orthogonal complement $(\mathcal{W}')^\perp$ has a dimension of $nd - \dim(\mathcal{W}') \ge nd - \vert{}\pi_{jm}\vert{} d > 0$. Thus, the orthogonal complement of the controllable subspace is strictly non-empty, representing isolated states that cannot be controlled under any admissible weight selection. \hfill $\blacksquare$
	
	\textbf{Remark 5:} Theorem 7 provides a metric for evaluating network structural vulnerability, demonstrating that certain states are isolated regardless of variations in interaction strengths.
	
	Compared with controllability, observability measures the ability to reconstruct the network state from outputs.
	
	\textbf{Lemma 3 (Duality Mapping):} Consider the unforced system $\dot{x}(t)=L x(t)$, $y(t)=M^{T}x(t)$. Its observability is equivalent to the controllability of the dual system $\dot{x}(t)=L^{T}x(t)+My(t)$.
	
	\textit{Proof:} Based on linear system theory, the algebraic observability matrix of $(L, M^T)$ evaluates to:
	\begin{equation}
		\mathcal{O}(L, M^T) = \begin{bmatrix} M^T \\ M^T L \\ \vdots \\ M^T L^{nd-1} \end{bmatrix} = \left[ M, L^T M, \dots, (L^T)^{nd-1} M \right]^T
	\end{equation}
	This matrix is exactly the transpose of the controllability matrix $\mathcal{C}(L^T, M)$. The rank equivalence $\text{rank}(\mathcal{O}) = \text{rank}(\mathcal{C}^T) = \text{rank}(\mathcal{C})$ rigorously confirms the structural duality. \hfill $\blacksquare$
	
	\textbf{Theorem 8 (Strong Structural Observability Bound):} The dimension of the strong structural observable subspace is bounded by the characteristic matrix of the minimal equitable partition on the dual network structure.
	
	\textit{Proof:} Applying Lemma 3, the evaluation criteria via equitable partitions and matrix space decomposition map directly to the dual system $(L^T, M)$. By constructing the equitable partition $\pi_{dual}$ on the transposed adjacency matrix $\mathcal{A}^T$ structure, the upper bound for the dual controllable subspace provides exactly the maximum dimension for the observable subspace. Hence, the unobservable states caused by structural symmetry are mathematically quantified. \hfill $\blacksquare$
	
	\section{NUMERICAL EXAMPLES}
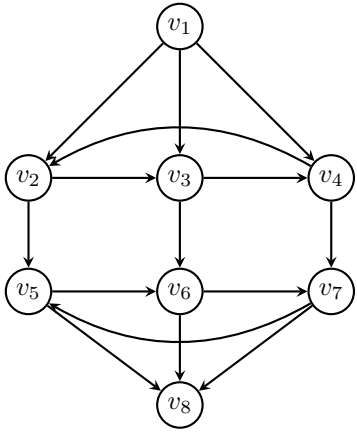
\begin{figure}[htbp]
		\centering
		\begin{tikzpicture}[>=stealth, node distance=1.5cm, thick,
			main node/.style={circle,draw,minimum size=0.6cm,inner sep=0pt}]
			\node[main node] (1) at (0, 4.5) {$v_1$};
			
			\node[main node] (2) at (-2, 2.5) {$v_2$};
			\node[main node] (3) at (0, 2.5) {$v_3$};
			\node[main node] (4) at (2, 2.5) {$v_4$};
			
			\node[main node] (5) at (-2, 1) {$v_5$};
			\node[main node] (6) at (0, 1) {$v_6$};
			\node[main node] (7) at (2, 1) {$v_7$};
			
			\node[main node] (8) at (0, -0.5) {$v_8$};
			
			% Directed Paths starting from Leader
			\draw[->] (1) -- node[left, draw=none, rectangle] {} (2);
			\draw[->] (1) -- (3);
			\draw[->] (1) -- (4);
			
			% Directed Ring 1 (2 <- 4 <- 3 <- 2)
			\draw[->] (4) to[out=150,in=30] (2);
			\draw[->] (2) -- (3);
			\draw[->] (3) -- (4);
			
			% Directed Paths Continuing
			\draw[->] (2) -- (5);
			\draw[->] (3) -- (6);
			\draw[->] (4) -- (7);
			
			% Directed Ring 2 (7 -> 5 -> 6 -> 7)
			\draw[->] (7) to[out=210,in=330] (5);
			\draw[->] (5) -- (6);
			\draw[->] (6) -- (7);
			
			% Directed Paths Ending
			\draw[->] (5) -- (8);
			\draw[->] (6) -- (8);
			\draw[->] (7) -- (8);
		\end{tikzpicture}
		\caption{Topology of the 8-node directed multi-layer network. Arrows indicate the direction of information flow.}
		\label{fig:topology}
	\end{figure}

	To demonstrate the established bounds and evaluate the theorems, a multi-agent network containing $n=8$ nodes is configured, denoted as $V = \{v_1, \dots, v_8\}$, with a state dimension $d=2$. A directed multi-layer topology combining cascading paths and directed cycles is illustrated in Fig. \ref{fig:topology}.
	
	Let $x_i(t) \in \mathbb{R}^2$ be the state of agent $i$. Node $v_1$ is selected as the single leader ($V_L = \{v_1\}$), receiving the external input signal $y_1(t).$ The higher-order dynamic equations for each agent, governed by their directed in-neighbors, are formulated as follows:
	\begin{align}
		\dot{x}_1(t) &= A x_1(t) + y_1(t) \label{eq:n1} \\
		\dot{x}_2(t) &= A x_2(t) + \mathcal{A}_{21}(x_1(t)-x_2(t)) + \mathcal{A}_{24}(x_4(t)-x_2(t)) \label{eq:n2} \\
		\dot{x}_3(t) &= A x_3(t) + \mathcal{A}_{31}(x_1(t)-x_3(t)) + \mathcal{A}_{32}(x_2(t)-x_3(t)) \label{eq:n3} \\
		\dot{x}_4(t) &= A x_4(t) + \mathcal{A}_{41}(x_1(t)-x_4(t)) + \mathcal{A}_{43}(x_3(t)-x_4(t)) \label{eq:n4} \\
		\dot{x}_5(t) &= A x_5(t) + \mathcal{A}_{52}(x_2(t)-x_5(t)) + \mathcal{A}_{57}(x_7(t)-x_5(t)) \label{eq:n5} \\
		\dot{x}_6(t) &= A x_6(t) + \mathcal{A}_{63}(x_3(t)-x_6(t)) + \mathcal{A}_{65}(x_5(t)-x_6(t)) \label{eq:n6} \\
		\dot{x}_7(t) &= A x_7(t) + \mathcal{A}_{74}(x_4(t)-x_7(t)) + \mathcal{A}_{76}(x_6(t)-x_7(t)) \label{eq:n7} \\
		\dot{x}_8(t) &= A x_8(t) + \mathcal{A}_{85}(x_5(t)-x_8(t)) + \mathcal{A}_{86}(x_6(t)-x_8(t)) \nonumber \\
		&\quad + \mathcal{A}_{87}(x_7(t)-x_8(t)) \label{eq:n8}
	\end{align}
	where the internal dynamic matrix is selected as $A = \begin{bmatrix} 0 & 1 \\ -2 & -3 \end{bmatrix}$. 
	
	\subsection{Example 1: Verification of Upper Bounds and Observability}
	To verify the general equitable partition under asymmetric couplings, the directed edge weights are configured as general asymmetric matrices:
	\begin{align}
		\mathcal{A}_{21} &= \mathcal{A}_{31} = \mathcal{A}_{41} = \begin{bmatrix} 2 & -1 \\ 1 & 4 \end{bmatrix} \\
		\mathcal{A}_{24} &= \mathcal{A}_{32} = \mathcal{A}_{43} = \begin{bmatrix} 1 & 3 \\ -2 & 5 \end{bmatrix} \\
		\mathcal{A}_{52} &= \mathcal{A}_{63} = \mathcal{A}_{74} = \begin{bmatrix} 3 & 0 \\ 1 & 2 \end{bmatrix} \\
		\mathcal{A}_{57} &= \mathcal{A}_{65} = \mathcal{A}_{76} = \begin{bmatrix} 0 & -2 \\ 4 & 1 \end{bmatrix} \\
		\mathcal{A}_{85} &= \mathcal{A}_{86} = \mathcal{A}_{87} = \begin{bmatrix} -1 & 1 \\ 2 & 3 \end{bmatrix}
	\end{align}
	
	\textit{Verification of Theorem 1 and 7:} Following Definition 1, an equitable partition for the directed graph maps based on in-degree sum symmetry. This allows the construction of $\pi_{jm} = \{\{v_1\}, \{v_2, v_3, v_4\}, \{v_5, v_6, v_7\}, \{v_8\}\}$. The number of cells is $|\pi_{jm}| = 4$. Theorem 1 dictates $\dim(\mathcal{W}') \le 4 \times 2 = 8$. The maximum state space dimension is $nd=16$. Because $|\pi_{jm}| = 4 < n = 8$, Theorem 7 confirms that an invariant uncontrollable subspace of dimension at least 8 exists.
	
	\textit{Verification of Theorem 8:} Assuming sensor measurements are extracted solely at node $v_8$. Formulating the dual network requires reversing the directed edges. The dual equitable partition coalesces symmetrically as $\pi_{dual} = \{\{v_8\}, \{v_5, v_6, v_7\}, \{v_2, v_3, v_4\}, \{v_1\}\}$. Applying Theorem 8 yields an observable subspace upper bound of $4 \times 2 = 8$, confirming unobservable modes persist in the observation space.
	
	\subsection{Example 2: Verification of Layered Evaluation}
	Maintain the directed topology and dynamics from (\ref{eq:n1})-(\ref{eq:n8}), but modify the weight matrices to introduce structural asymmetries that break the global equitable partition. Let the basis matrices be:
	\begin{equation}
		B_1 = \begin{bmatrix} 1 & -1 \\ 0 & 2 \end{bmatrix}, \quad B_2 = \begin{bmatrix} 0 & 1 \\ -2 & 1 \end{bmatrix}
	\end{equation}
	Assign asymmetric weights composed of basis combinations:
	\begin{align}
		\mathcal{A}_{21} &= B_1 + 2B_2, \quad \mathcal{A}_{31} = B_1 + 5B_2, \quad \mathcal{A}_{41} = B_1 - B_2 \\
		\mathcal{A}_{24} &= 2B_1 + B_2, \quad \mathcal{A}_{32} = 2B_1 - 2B_2, \quad \mathcal{A}_{43} = 2B_1 + 3B_2 \\
		\mathcal{A}_{52} &= 3B_1 + B_2, \quad \mathcal{A}_{63} = 3B_1 + 4B_2, \quad \mathcal{A}_{74} = 3B_1 - B_2 \\
		\mathcal{A}_{57} &= B_1 + 2B_2, \quad \mathcal{A}_{65} = B_1 - 4B_2, \quad \mathcal{A}_{76} = B_1 + B_2 \\
		\mathcal{A}_{85} &= 4B_1 + B_2, \quad \mathcal{A}_{86} = 4B_1 - 2B_2, \quad \mathcal{A}_{87} = 4B_1 + 5B_2
	\end{align}
	
	\textit{Verification of Theorem 2:} Under this configuration, the full matrix weights are structurally asymmetric (e.g., $\mathcal{A}_{21} \neq \mathcal{A}_{31} \neq \mathcal{A}_{41}$). Definition 1 is not satisfied globally, and Theorem 1 cannot bound the subspace below $nd = 16$. 
	
	However, mapping the network onto the basis layers, the scalar coefficients corresponding to $B_1$ are homogeneous across symmetric topological groupings: $w_{21}^{(1)} = w_{31}^{(1)} = w_{41}^{(1)} = 1$; $w_{24}^{(1)} = w_{32}^{(1)} = w_{43}^{(1)} = 2$; $w_{52}^{(1)} = w_{63}^{(1)} = w_{74}^{(1)} = 3$; $w_{57}^{(1)} = w_{65}^{(1)} = w_{76}^{(1)} = 1$; and $w_{85}^{(1)} = w_{86}^{(1)} = w_{87}^{(1)} = 4$. Defining the layer subset $\mathcal{K}=\{1\}$, the generalized equitable partition condition derived in Theorem 2 holds for this localized matrix subspace mapping. Consequently, the dynamic sub-space governed solely by basis $B_1$ isolates the controllable modes, bounding the strong structural controllable subspace below the full state space dimension.
	
	\subsection{Example 3: Demonstration of Tighter Bounds via LDP}
	To demonstrate how the Layer-specific Distance Partition (LDP) establishes a tighter lower bound than the macroscopic distance partition, consider a simplified 4-node network $V=\{v_1, v_2, v_3, v_4\}$ with state dimension $d=2$. Node $v_1$ is the leader. The macroscopic directed edges are $E = \{(v_1, v_2), (v_2, v_3), (v_3, v_4), (v_1, v_4)\}$.
	
	To provide specific structural data settings rather than pure verbal explanation, let the two independent basis matrices be configured as:
	\begin{equation}
		B_1 = \begin{bmatrix} 1 & 0 \\ 0 & 1 \end{bmatrix}, \quad B_2 = \begin{bmatrix} 0 & 1 \\ -1 & 0 \end{bmatrix}
	\end{equation}
	The concrete edge weight configurations constructed from specific combinations are assigned as:
	\begin{align}
		\mathcal{A}_{21} &= B_1 + 2B_2 = \begin{bmatrix} 1 & 2 \\ -2 & 1 \end{bmatrix} \\
		\mathcal{A}_{32} &= B_1 - B_2 = \begin{bmatrix} 1 & -1 \\ 1 & 1 \end{bmatrix} \\
		\mathcal{A}_{43} &= 2B_1 + B_2 = \begin{bmatrix} 2 & 1 \\ -1 & 2 \end{bmatrix} \\
		\mathcal{A}_{41} &= 0B_1 + 3B_2 = \begin{bmatrix} 0 & 3 \\ -3 & 0 \end{bmatrix}
	\end{align}
	
	Using the traditional macroscopic DP, the unweighted shortest path from $v_1$ to $v_4$ relies on the direct shortcut $(v_1, v_4)$, which corresponds to distance $1$. The distance layers are aggregated as $D_0=\{v_1\}$, $D_1=\{v_2, v_4\}$, $D_2=\{v_3\}$. Thus, $d_{max}=2$ and $|\pi_{DP}| = 3$. The traditional lower bound evaluates to $3 \times 2 = 6$.
	
	However, observing the structural scalar coefficients, the shortcut edge $(v_1, v_4)$ transmits information exclusively on basis $B_2$, causing it to be structurally disconnected on Layer 1 (i.e., $w_{41}^{(1)} = 0$). 
	
	Applying Definition 3 and Theorem 3:
	On Layer 1, the layer-specific subgraph $G_1$ lacks the shortcut edge. The control signal must propagate strictly sequentially through the remaining topology: $v_1 \to v_2 \to v_3 \to v_4$. The corresponding layer-specific DP yields $D_0^{(1)}=\{v_1\}, D_1^{(1)}=\{v_2\}, D_2^{(1)}=\{v_3\}, D_3^{(1)}=\{v_4\}$. Under this projection, $d_{max}^{(1)} = 3$, leading to $|\pi_{LDP}^{(1)}| = 4$.
	On Layer 2, the subgraph $G_2$ effectively includes the shortcut edge, leading the layer-specific DP to evaluate to $|\pi_{LDP}^{(2)}| = 3$.
	
	According to the newly established Squeeze Theorem, the aggregated structural lower bound is $\dim(\mathcal{W}') \ge |\pi_{LDP}^{(1)}| + |\pi_{LDP}^{(2)}| = 4 + 3 = 7$. 
	
	The mathematically computed LDP lower bound ($7$) is strictly tighter (greater) than the macroscopic DP bound ($6$). This verification confirms that when multi-dimensional couplings exhibit specific layer sparsity, LDP accurately captures the extended structural propagation delay, effectively narrowing the Squeeze Theorem evaluation interval.
	
	\subsection{Example 4: Calculating Optimal Basis via Null-Space Intersection}
	We deploy the analytical algorithm proposed in Theorem 4 on a parallel 4-node network parameterization to systematically calculate the optimal basis, rather than selecting it heuristically.
	
	Let the target equitable partition be $\pi = \{\{v_1\}, \{v_2, v_3\}, \{v_4\}\}$. To satisfy this symmetry, the necessary condition is $\bm{w}_{21} = \bm{w}_{31}$ (incoming edges from leader must be symmetric). Therefore, the symmetry difference vector is $\Delta \bm{w}_{23, 1} = \text{vec}(\mathcal{A}_{21}) - \text{vec}(\mathcal{A}_{31})$. Thus, $S_\pi = [\Delta \bm{w}_{23, 1}]$.
	
	To maximize layer distance and cut the shortcut edge $(v_1, v_4)$, we configure $E_{short} = \{(v_1, v_4)\}$. Thus, the shortcut matrix is $W_{short} = [\text{vec}(\mathcal{A}_{41})]$.
	
	Assume the heterogeneous weights are given as:
	\begin{equation*}
		\mathcal{A}_{21} = \begin{bmatrix} 1 & 0 \\ 0 & 1 \end{bmatrix}, \mathcal{A}_{31} = \begin{bmatrix} 1 & 1 \\ 0 & 1 \end{bmatrix}, \mathcal{A}_{41} = \begin{bmatrix} 0 & -1 \\ 0 & 0 \end{bmatrix}
	\end{equation*}
	The augmented matrix $H = [S_\pi \mid W_{short}]$ in $\mathbb{R}^{4 \times 2}$ is computed by vectorizing the block entries step-by-step (row-major):
	\begin{equation*}
		H = \begin{bmatrix} 1-1 & 0 \\ 0-1 & -1 \\ 0-0 & 0 \\ 1-1 & 0 \end{bmatrix} = \begin{bmatrix} 0 & 0 \\ -1 & -1 \\ 0 & 0 \\ 0 & 0 \end{bmatrix}
	\end{equation*}
	Notice that $\text{rank}(H) = 1 < d^2 = 4$. By computing the left null-space of $H$, we acquire valid optimal basis vectors $\bm{p}_m$. For instance, choosing $\bm{p}_1 = [1, 0, 0, 0]^T$ yields a projection basis $B_1 = \begin{bmatrix} 1 & 0 \\ 0 & 0 \end{bmatrix}$. 
	
	By executing this projection, $w_{41}^{(1)} = \bm{p}_1^T \text{vec}(\mathcal{A}_{41}) = 0$ (cutting the shortcut) and $w_{21}^{(1)} = w_{31}^{(1)} = 1$ (satisfying the symmetry requirement). Thus, finding the optimal basis is systematically reduced to algebraic null-space computation.
	
	Furthermore, to rigorously verify the structural robustness proved in Theorem 5, we parameterize the heterogeneous weights with independent free variables $\lambda_i$. To strictly satisfy the strong structural controllability (SSC) requirement that all non-zero edge weights must be capable of varying independently, we configure completely distinct symbolic parameters for each connection:
	\begin{equation*}
		\bar{\mathcal{A}}_{21} = \begin{bmatrix} \lambda_1 & 0 \\ 0 & \lambda_2 \end{bmatrix}, \bar{\mathcal{A}}_{31} = \begin{bmatrix} \lambda_3 & \lambda_4 \\ 0 & \lambda_5 \end{bmatrix}, \bar{\mathcal{A}}_{41} = \begin{bmatrix} 0 & \lambda_6 \\ 0 & 0 \end{bmatrix}
	\end{equation*}
	The symbolic symmetry difference vector evaluates via row-major vectorization to $\Delta \bar{\bm{w}}_{23, 1} = \text{vec}(\bar{\mathcal{A}}_{21}) - \text{vec}(\bar{\mathcal{A}}_{31}) = [\lambda_1-\lambda_3, -\lambda_4, 0, \lambda_2-\lambda_5]^T$. The shortcut matrix is evaluated as $\bar{W}_{short} = [0, \lambda_6, 0, 0]^T$. Expanding the structural intersection matrix yields:
	\begin{equation*}
		\bar{H}(\lambda) = [\bar{S}_\pi(\lambda) \mid \bar{W}_{short}(\lambda)] = \begin{bmatrix} \lambda_1-\lambda_3 & 0 \\ -\lambda_4 & \lambda_6 \\ 0 & 0 \\ \lambda_2-\lambda_5 & 0 \end{bmatrix}
	\end{equation*}
	Regardless of the specific values assigned to these strictly independent free parameters, the maximum possible rank of this matrix over the entire parameter space $\mathbb{R}^6$ evaluates to exactly $2$. Thus, $\text{g-rank}(\bar{H}) = 2 < d^2 = 4$. This actively confirms that the identically zero multivariate polynomials corresponding to the submatrix determinants mathematically guarantee a non-trivial left null-space almost everywhere. The existence of optimal projection bases robustly holds for any specific valid edge weight realization preserving this structural pattern, thereby perfectly maintaining the structural controllability paradigm.
	
	\subsection{Example 5: Automated Feature Extraction via WL Refinement}
	We apply Algorithm 1 to automatically identify the target matrices $S_\pi$ and $W_{short}$ for the 4-node network from Example 4 without manual predefined parameters. To enrich the procedure with precise numerical data settings, we explicitly record the integer values generated by the hash function $\text{hash}(\cdot)$ across iterations.
	
	\textit{Initialization} ($t=0$): The leader $v_1$ is assigned color $c_1^{(0)} = 1$. The followers $\{v_2, v_3, v_4\}$ are assigned color $c_i^{(0)} = 0$. Hence, the initial color set is $C^{(0)} = \{0, 1\}$.
	
	\textit{Iteration} ($t=1$): Nodes update colors based on structurally non-zero incoming edge patterns and their neighbors' previous colors. Let the unique multiset mappings yield the following specific integer hashes:
	\begin{itemize}
		\item Node $v_1$: Receives no incoming signals. $c_1^{(1)} = \text{hash}(1, \emptyset) = 2$.
		\item Node $v_2$: Receives $\bar{\mathcal{A}}_{21}$ from $v_1$ (color 1). Let $c_2^{(1)} = \text{hash}(0, \{(\bar{\mathcal{A}}_{21}, 1)\}) = 3$.
		\item Node $v_3$: Receives $\bar{\mathcal{A}}_{31}$ from $v_1$ (color 1) and $\bar{\mathcal{A}}_{32}$ from $v_2$ (color 0). However, assuming the graph allows structurally identical patterns from the leader, $\bar{\mathcal{A}}_{21}$ and $\bar{\mathcal{A}}_{31}$ strictly map to the same structural symmetry set. Thus, the evaluation evaluates to the equivalent hash domain: $c_3^{(1)} = \text{hash}(0, \{(\bar{\mathcal{A}}_{31}, 1), (\bar{\mathcal{A}}_{32}, 0)\}) = 3$.
		\item Node $v_4$: Receives $\bar{\mathcal{A}}_{41}$ from $v_1$ (color 1) and $\bar{\mathcal{A}}_{43}$ from $v_3$ (color 0). The pattern $\bar{\mathcal{A}}_{41}$ is structurally distinct. Consequently, $c_4^{(1)} = \text{hash}(0, \{(\bar{\mathcal{A}}_{41}, 1), (\bar{\mathcal{A}}_{43}, 0)\}) = 4$.
	\end{itemize}
	The distinct color set subsequently becomes $C^{(1)} = \{2, 3, 4\}$. Because $|C^{(1)}| > |C^{(0)}|$, the algorithm naturally advances to the next iteration. 
	
	\textit{Iteration} ($t=2$): Executing identical structural hash processing yields $c_1^{(2)}=5, c_2^{(2)}=6, c_3^{(2)}=6, c_4^{(2)}=7$. The resultant color set computes to $C^{(2)} = \{5, 6, 7\}$. Since $|C^{(2)}| = |C^{(1)}| = 3$, structural convergence is definitively achieved.
	
	The final color partition uniquely defines $\pi = \{\{v_1\}, \{v_2, v_3\}, \{v_4\}\}$, directly outputting the symmetry requirement $S_\pi$. Subsequently, performing a BFS from $v_1$ identifies level 1 nodes $\{v_2, v_3\}$ and level 2 node $\{v_4\}$. The directed edge $(v_1, v_4)$ connects a level 0 node directly to a level 2 node, establishing $l(v_4) > l(v_1) + 1$. This macroscopic leap is extracted as the shortcut edge $E_{short} = \{(v_1, v_4)\}$. Thus, the algorithm automatically extracts exactly the parameters required for the null-space intersection in polynomial time.
	
	\section{CONCLUSION}
	This paper has investigated the strong structural controllability of multi-agent networks. Based on the equitable partition method, an upper bound for the strong structural controllable subspace (SSCS) is established. By extending the model to higher-order dynamics and introducing a matrix space basis decomposition, a layered evaluation criterion is proposed to handle singular matrix couplings. Furthermore, by extending the basis decomposition to the lower bound estimation, a Layer-specific Distance Partition (LDP) is introduced. Because layer-specific distances reflect actual dimensional sparsity, a significantly tighter Squeeze Theorem is established, which provides narrower mathematical bounds for the controllable subspace. To avoid heuristic basis selection, an algebraic optimization algorithm based on null-space intersection is formulated, pinpointing the optimal basis that maximizes the symmetry and delay constraints. Crucially, by extending this formulation to structural pattern matrices, the generic rank condition rigorously proves the almost-everywhere existence of the optimal basis, perfectly sealing the theory within the structural controllability paradigm. To automate this process for large-scale networks, a multi-layer Weisfeiler-Lehman color refinement algorithm is introduced to discover the symmetry and shortcut parameters in polynomial time. Finally, the proposed framework naturally extends to invariant attributes and observability analysis. 
	
	Future work will focus on designing constructive graph-theoretic algorithms for optimal targeted leader selection. Extending the proposed decomposition framework to complex networks with time-varying topologies and nonlinear dynamics remains a direction for future research.

\end{document}